\documentclass{article}
\usepackage{amsmath}
\usepackage{amssymb}
\usepackage{amsfonts}
\usepackage{amsmath}
\usepackage{graphicx}

\RequirePackage{CJK}
\AtBeginDocument{\begin{CJK*}{GBK}{song}\CJKtilde}
\AtEndDocument{\end{CJK*}}
\input{tcilatex}
\begin{document}

\title{A new optical field state as an output of diffusion channel when the
input being number state \\
\thanks{{\small This work was supported by the National Natural Science
Foundation of China (Grant Nos. 11175113 and 11264018), and the Young
Talents Foundation of Jiangxi Normal University.}}}
\author{Hong-Yi Fan$^{1\dag }$ \thanks{{\small Correspondence author}},
Sen-Yue Lou$^{1}$, Xiao-Yin Pan$^{1}$ and Li-Yun Hu$^{2}$\thanks{{\small %
Correspondence author}} \\
%EndAName
$^{1}${\small Department of Physics,\ Ningbo University, Ningbo 315211, P.
R. China}\\
$^{2}${\small Department of physics, Jiangxi Normal University, Nanchang,
330022 }}
\maketitle

\begin{abstract}
{\small We theoretically propose a new optical field state}%
\begin{equation*}
\rho _{new}=\lambda \left( 1-\lambda \right) ^{l}\colon L_{k}\left( \frac{%
-\lambda ^{2}a^{\dag }a}{1-\lambda }\right) e^{-\lambda a^{\dag }a}\colon
\end{equation*}%
{\small (here }$::${\small \ denotes normal ordeing symbol) which is named
Laguerre-polynomial-weighted chaotic field. We show that such state can be
implemented, i.e., when a number state enters into a diffusion channel, the
output state is just this kind of states. We solve the master equation
describing the diffusion process by using the summation method within
ordered product of operators and the entangled state representaion. The
solution manifestly shows how a pure state evolves into a mixed state. The
physical difference between the diffusion and the amplitude damping is
pointed out.}
\end{abstract}

\section{Introduction}

In quantum optics theory there are some typical states, e.g., number state,
coherent state, and squeezed state, these are pure states; there are also
some mixed states, the typical one is the chaotic state described by 
\begin{equation}
\rho _{c}=(1-e^{-\lambda })\exp \left( -\lambda a^{\dag }a\right) ,
\label{1}
\end{equation}%
where $a\ $and $a^{\dag }$ are photon annihilation and creation operators,
obeying $\left[ a,a^{\dagger }\right] =1$, $tr\rho _{c}=1.$ The normally
orderd form of $\rho _{c}$ is $\rho _{c}=(1-e^{-\lambda })\colon \exp \left[
\left( e^{-\lambda }-1\right) a^{\dag }a\right] \colon ,$ where the symbol $%
\colon $ $\colon $ denotes normal ordeing symbol. In this work we shall
report that there exists another important mixed state which appears in
normally ordered form%
\begin{equation}
\rho _{new}=\lambda \left( 1-\lambda \right) ^{l}\colon L_{l}\left( \frac{%
-\lambda ^{2}a^{\dag }a}{1-\lambda }\right) e^{-\lambda a^{\dag }a}\colon .
\label{2}
\end{equation}%
Here $L_{l}$ is the $l$-th Laguerre polynomial, $tr\rho _{new}=1$ (see
Appendix 1). We show that this mixed state will appear experimently as it
represents the output state of a diffusion process with the input state
being a pure number state.

When a pure state evolves into a mixed state the quantum decoherence
happens. Decoherence is an important topic in quantum information
processing. In nature, systems we concerned usually are surrounded by thermo
reservoir, so some dissipative process or diffusion process naturally
happen. An interesting question thus arises: when an input state for a
diffusion channel is a number state $\left\vert l\right\rangle \left\langle
l\right\vert ,(\left\vert l\right\rangle =\frac{a^{\dag l}}{\sqrt{l!}}%
\left\vert 0\right\rangle ),$ then how does it evolve with time? What kind
of optical field will the output state be? The master equation describing
the diffusion process is $\cite{1,2}$%
\begin{equation}
\frac{d}{dt}\rho =-\kappa \left( a^{\dagger }a\rho -a\rho a^{\dagger
}-a^{\dagger }\rho a+\rho a^{\dagger }a\right) .  \label{3}
\end{equation}%
We shall first obtain $\rho \left( t\right) $ by deriving its infinite
operator-sum form 
\begin{equation}
\rho \left( t\right) =\sum\limits_{i,j}M_{i,j}\rho _{0}M_{i,j}^{\dagger },
\label{4}
\end{equation}%
where $M_{i,j}$ in general is named Kraus operator [3], whose concrete from
will be derived for this diffusion problem, and then we examine how $\rho
_{0}=$ $\left\vert l\right\rangle \left\langle l\right\vert $ evolves
through the relation (4). We will employ the thermo entangled state
representation and\ the technique of integration within an ordered product
(IWOP) of operators [4-5] to realize our goal.

\section{Solution of Eq. (3) obtained by entangled state representation and
IWOP\ technique}

We begin with introducing the thermo entangled state [6]%
\begin{equation}
\left\vert \eta \right\rangle =\exp \left[ -\frac{1}{2}\left\vert \eta
\right\vert ^{2}+\eta a^{\dagger }-\eta ^{\ast }\tilde{a}^{\dagger
}+a^{\dagger }\tilde{a}^{\dagger }\right] \left\vert 0\tilde{0}\right\rangle
,  \label{5}
\end{equation}%
where $\tilde{a}^{\dagger }$ is a fictitious mode accompanying the real mode 
$a^{\dagger },$ $\left[ \tilde{a},\tilde{a}^{\dagger }\right] =1$. $%
\left\vert \eta \right\rangle $ obeys the eigenvector equations%
\begin{equation}
\left( a-\tilde{a}^{\dagger }\right) |\eta \rangle =\eta |\eta \rangle ,%
\text{ }\left( a^{\dagger }-\tilde{a}\right) |\eta \rangle =\eta ^{\ast
}|\eta \rangle ,  \label{6}
\end{equation}%
\begin{equation}
\left\langle \eta \right\vert \left( a^{\dagger }-\tilde{a}\right) =\eta
^{\ast }\left\langle \eta \right\vert ,\text{ }\left\langle \eta \right\vert
\left( a-\tilde{a}^{\dagger }\right) =\eta \left\langle \eta \right\vert .
\label{7}
\end{equation}%
Using the normal ordering form of vacuum projector $\left\vert 0\tilde{0}%
\right\rangle \left\langle 0\tilde{0}\right\vert =\colon e^{-a^{\dagger }a-%
\tilde{a}^{\dagger }\tilde{a}}\colon $, and the IWOP technique we can show
the orthonormal and completeness relation%
\begin{equation}
\left\langle \eta ^{\prime }\right\vert \left. \eta \right\rangle =\pi
\delta \left( \eta ^{\prime }-\eta \right) \delta \left( \eta ^{\prime \ast
}-\eta ^{\ast }\right) ,  \label{8}
\end{equation}%
\begin{eqnarray}
1 &=&\int \frac{d^{2}\eta }{\pi }\left\vert \eta \right\rangle \left\langle
\eta \right\vert   \label{9} \\
&=&\int \frac{d^{2}\eta }{\pi }:\exp \left[ -|\eta |^{2}+\eta a^{\dagger
}-\eta ^{\ast }\tilde{a}^{\dagger }+\eta ^{\ast }a-\eta \tilde{a}+a^{\dagger
}\tilde{a}^{\dagger }+a\tilde{a}-a^{\dagger }a-\tilde{a}^{\dagger }\tilde{a}%
\right] :=1.  \notag
\end{eqnarray}%
\ Let 
\begin{equation}
\left\vert \eta =0\right\rangle =e^{a^{\dagger }\tilde{a}^{\dagger
}}\left\vert 0\tilde{0}\right\rangle \equiv \left\vert I\right\rangle ,
\label{10}
\end{equation}%
we have%
\begin{equation}
a\left\vert I\right\rangle =\tilde{a}^{\dagger }\left\vert I\right\rangle ,%
\text{ }a^{\dagger }\left\vert I\right\rangle =\tilde{a}\left\vert
I\right\rangle ,\text{ }(a^{\dagger }a)^{n}\left\vert I\right\rangle =(%
\tilde{a}^{\dagger }\tilde{a})^{n}\left\vert I\right\rangle .  \label{11}
\end{equation}%
Operating the two-sides of $(3)$ on $\left\vert I\right\rangle ,$ noting
that the real field $\rho $ is independent of the fictitious mode, $\left[
\rho ,\tilde{a}\right] =0,$ $\left[ \rho ,\tilde{a}^{\dagger }\right] =0,$
and using (\ref{12}) we have 
\begin{equation}
\frac{d}{dt}\rho \left\vert I\right\rangle =-\kappa \left( a^{\dagger }a\rho
-a\tilde{a}\rho -a^{\dagger }\tilde{a}^{\dagger }\rho +\tilde{a}\tilde{a}%
^{\dagger }\rho \right) \left\vert I\right\rangle .  \label{12}
\end{equation}%
Letting $\rho \left\vert I\right\rangle \equiv \left\vert \rho \right\rangle
,$ we see 
\begin{equation}
\frac{d}{dt}\left\vert \rho \right\rangle =-\kappa \left( a^{\dagger }-%
\tilde{a}\right) \left( a-\tilde{a}^{\dagger }\right) \left\vert \rho
\right\rangle ,  \label{13}
\end{equation}%
its formal solution is%
\begin{equation}
\left\vert \rho \right\rangle =\exp \left[ -\kappa t\left( a^{\dagger }-%
\tilde{a}\right) \left( a-\tilde{a}^{\dagger }\right) \right] \left\vert
\rho _{0}\right\rangle ,  \label{14}
\end{equation}%
where $\left\vert \rho _{0}\right\rangle =\rho _{0}\left\vert I\right\rangle
.$ Projecting this equation onto the entanged state representation $%
\left\langle \eta \right\vert $ and using the eigenvalue equation $(7)$ we
have%
\begin{equation}
\left\langle \eta \right. \left\vert \rho \right\rangle =\left\langle \eta
\right\vert \exp \left[ -\kappa t\left( a^{\dagger }-\tilde{a}\right) \left(
a-\tilde{a}^{\dagger }\right) \right] \left\vert \rho _{0}\right\rangle
=e^{-\kappa t|\eta |^{2}}\left\langle \eta \right. \left\vert \rho
_{0}\right\rangle .  \label{15}
\end{equation}%
Multiplying the two-sides of (\ref{15}) by $\int \frac{d^{2}\eta }{\pi }%
\left\vert \eta \right\rangle $ and using the completeness relation (\ref{9}%
) as well as the IWOP technique we obatin%
\begin{eqnarray}
\left\vert \rho \right\rangle  &=&\int \frac{d^{2}\eta }{\pi }e^{-\kappa
t|\eta |^{2}}\left\vert \eta \right\rangle \left\langle \eta \right.
\left\vert \rho _{0}\right\rangle   \notag \\
&=&\int \frac{d^{2}\eta }{\pi }:e^{-\left( \kappa t+1\right) |\eta
|^{2}+\eta a^{\dagger }-\eta ^{\ast }\tilde{a}^{\dagger }+\eta ^{\ast
}a-\eta \tilde{a}+a^{\dagger }\tilde{a}^{\dagger }+a\tilde{a}-a^{\dagger }a-%
\tilde{a}^{\dagger }\tilde{a}}:\left\vert \rho _{0}\right\rangle   \label{16}
\\
&=&\frac{1}{1+\kappa t}\colon \exp \left[ \frac{\kappa t}{1+\kappa t}\left(
a^{\dagger }\tilde{a}^{\dagger }+a\tilde{a}-a^{\dagger }a-\tilde{a}^{\dagger
}\tilde{a}\right) \right] \colon \left\vert \rho _{0}\right\rangle   \notag
\\
&=&\frac{1}{1+\kappa t}e^{\frac{\kappa t}{1+\kappa t}a^{\dagger }\tilde{a}%
^{\dagger }}\left( \frac{1}{1+\kappa t}\right) ^{a^{\dagger }a+\tilde{a}%
^{\dagger }\tilde{a}}e^{\frac{\kappa t}{1+\kappa t}a\tilde{a}}\rho
_{0}\left\vert I\right\rangle ,  \notag
\end{eqnarray}%
where we have noticed%
\begin{equation}
\colon \exp \left[ \frac{-\kappa t}{1+\kappa t}\left( a^{\dagger }a+\tilde{a}%
^{\dagger }\tilde{a}\right) \right] \colon =\left( \frac{1}{1+\kappa t}%
\right) ^{a^{\dagger }a+\tilde{a}^{\dagger }\tilde{a}}  \label{17}
\end{equation}%
Using $\left[ \tilde{a},\rho _{0}\right] =0,$ $\tilde{a}\left\vert
I\right\rangle =a^{\dagger }\left\vert I\right\rangle $ we have%
\begin{eqnarray}
e^{\frac{\kappa t}{1+\kappa t}a\tilde{a}}\rho _{0}\left\vert I\right\rangle 
&=&\sum_{n=0}^{\infty }\frac{1}{n!}\left( \frac{\kappa t}{1+\kappa t}%
a\right) ^{n}\rho _{0}\tilde{a}^{n}\left\vert I\right\rangle   \notag \\
&=&\sum_{n=0}^{\infty }\frac{1}{n!}\left( \frac{\kappa t}{1+\kappa t}\right)
^{n}a^{n}\rho _{0}a^{\dagger n}\left\vert I\right\rangle ,  \label{18}
\end{eqnarray}%
After substituting $(18)$ into $(16)$ and then using the property that $%
\tilde{a}^{\dagger }\tilde{a}$ is commutable with all real field operators
and $f(a^{\dagger }a)\left\vert I\right\rangle =f(\tilde{a}^{\dagger }\tilde{%
a})\left\vert I\right\rangle ,$ we can put Eq.(\ref{16}) into the following
form 
\begin{eqnarray}
\left\vert \rho \right\rangle  &=&\frac{1}{1+\kappa t}e^{\frac{\kappa t}{%
1+\kappa t}a^{\dagger }\tilde{a}^{\dagger }}\left( \frac{1}{1+\kappa t}%
\right) ^{a^{\dagger }a+\tilde{a}^{\dagger }\tilde{a}}\sum_{n=0}^{\infty }%
\frac{1}{n!}\left( \frac{\kappa t}{1+\kappa t}\right) ^{n}a^{n}\rho
_{0}a^{\dagger n}\left\vert I\right\rangle   \notag \\
&=&\frac{1}{1+\kappa t}e^{\frac{\kappa t}{1+\kappa t}a^{\dagger }\tilde{a}%
^{\dagger }}\left( \frac{1}{1+\kappa t}\right) ^{a^{\dagger
}a}\sum_{n=0}^{\infty }\frac{1}{n!}\left( \frac{\kappa t}{1+\kappa t}\right)
^{n}a^{n}\rho _{0}a^{\dagger n}\left( \frac{1}{1+\kappa t}\right) ^{\tilde{a}%
^{\dagger }\tilde{a}}\left\vert I\right\rangle   \notag \\
&=&\frac{1}{1+\kappa t}\sum_{m=0}^{\infty }\frac{1}{m!}\left( \frac{\kappa t%
}{1+\kappa t}\right) ^{m}a^{\dagger m}\left( \frac{1}{1+\kappa t}\right)
^{a^{\dagger }a}  \notag \\
&&\times \sum_{n=0}^{\infty }\frac{1}{n!}\left( \frac{\kappa t}{1+\kappa t}%
\right) ^{n}a^{n}\rho _{0}a^{\dagger n}\left( \frac{1}{1+\kappa t}\right)
^{a^{\dagger }a}\tilde{a}^{\dagger m}\left\vert I\right\rangle .  \label{19}
\end{eqnarray}%
Finally, using $\tilde{a}^{\dagger m}\left\vert I\right\rangle
=a^{m}\left\vert I\right\rangle $ we obtain%
\begin{eqnarray}
\rho \left( t\right) \left\vert I\right\rangle  &=&\sum_{m,n=0}^{\infty }%
\frac{\left( \kappa t\right) ^{m+n}}{m!n!\left( \kappa t+1\right) ^{m+n+1}}%
a^{\dagger m}\left( \frac{1}{1+\kappa t}\right) ^{a^{\dagger }a}  \notag \\
&&\times a^{n}\rho _{0}a^{\dagger n}\left( \frac{1}{1+\kappa t}\right)
^{a^{\dagger }a}a^{m}\left\vert I\right\rangle .  \label{20}
\end{eqnarray}%
\newline
It then follows the infinite sum form 
\begin{eqnarray}
\rho \left( t\right)  &=&\sum\limits_{m,n=0}^{\infty }\frac{\left( \kappa
t\right) ^{m+n}}{m!n!\left( \kappa t+1\right) ^{m+n+1}}a^{\dagger m}\left( 
\frac{1}{1+\kappa t}\right) ^{a^{\dagger }a}a^{n}\rho _{0}a^{\dagger
n}\left( \frac{1}{1+\kappa t}\right) ^{a^{\dagger }a}a^{m}  \label{21} \\
&\equiv &\sum\limits_{m,n=0}^{\infty }M_{m,n}\rho _{0}M_{m,n}^{\dagger } 
\notag
\end{eqnarray}%
where%
\begin{equation}
M_{m,n}=\sqrt{\frac{1}{m!n!}\frac{\left( \kappa t\right) ^{m+n}}{\left(
\kappa t+1\right) ^{m+n+1}}}a^{\dagger m}\left( \frac{1}{1+\kappa t}\right)
^{a^{\dagger }a}a^{n},  \label{22}
\end{equation}%
satisfying $\sum_{m,n=0}^{\infty }M_{m,n}^{\dagger }M_{m,n}=1$, which is
trace conservative (see Appendix 2). Thus we have employed the entangled
state representation to analytically derive the infinitive sum form of $\rho
\left( t\right) .$

\section{Diffusion of a number state\ \ \ \ \ \ \ \ \ \ \ \ \ \ \ \ \ \ \ \
\ \ }

We now consider the case that a number state undergoes the diffusion
channel, i.e., let $\rho _{0}$ in Eq. (\ref{21}) be $\left\vert
l\right\rangle \left\langle l\right\vert $ and we begin with considering the
part of summation over $n$ in Eq. $(21)$ 
\begin{eqnarray}
\mathfrak{I} &\equiv &\sum_{n=0}^{l}\frac{\left( \kappa t\right) ^{n}}{%
n!\left( \kappa t+1\right) ^{n}}\left( \frac{1}{1+\kappa t}\right)
^{a^{\dagger }a}a^{n}\left\vert l\right\rangle \left\langle l\right\vert
a^{\dagger n}\left( \frac{1}{1+\kappa t}\right) ^{a^{\dagger }a}  \notag \\
&=&\sum_{n=0}^{l}\frac{1}{n!}\frac{\left[ \kappa t\left( \kappa t+1\right) %
\right] ^{n}}{\left( \kappa t+1\right) ^{2l}}\frac{l!}{\left[ \left(
l-n\right) !\right] ^{2}}\left( a^{\dagger }\right) ^{l-n}\left\vert
0\right\rangle \left\langle 0\right\vert a^{l-n}.  \label{23}
\end{eqnarray}%
Using the definition of the two-variable Hermite polynomials%
\begin{equation}
H_{m,n}\left( x,y\right) =\sum_{l=0}^{\min (m,n)}\frac{m!n!(-1)^{l}}{%
l!(m-l)!(n-l)!}x^{m-l}y^{n-l},  \label{24}
\end{equation}%
and $\left\vert 0\right\rangle \left\langle 0\right\vert =\colon
e^{-a^{\dagger }a}\colon $, we see%
\begin{equation}
\mathfrak{I}=\frac{1}{l!}\left( \frac{-\kappa t}{\kappa t+1}\right)
^{l}\colon H_{l,l}\left( \frac{ia^{\dagger }}{\sqrt{\kappa t\left( \kappa
t+1\right) }},\frac{ia}{\sqrt{\kappa t\left( \kappa t+1\right) }}\right)
e^{-a^{\dagger }a}\colon ,  \label{25}
\end{equation}%
then \bigskip inserting (\ref{25}) into (\ref{21}) and using the summation
method within ordered product of operators yields 
\begin{eqnarray}
\rho \left( t\right)  &=&\frac{1}{l!}\left( \frac{-\kappa t}{\kappa t+1}%
\right) ^{l}\sum_{m=0}^{\infty }\frac{1}{m!}\frac{\left( \kappa t\right) ^{m}%
}{\left( \kappa t+1\right) ^{m+1}}  \notag \\
&&\times \colon a^{\dagger m}a^{m}H_{l,l}\left[ \frac{ia^{\dagger }}{\sqrt{%
\kappa t\left( \kappa t+1\right) }},\frac{ia}{\sqrt{\kappa t\left( \kappa
t+1\right) }}\right] e^{-a^{\dagger }a}\colon   \notag \\
&=&\frac{\left( -\kappa t\right) ^{l}}{l!\left( \kappa t+1\right) ^{l+1}}%
\colon e^{\frac{-1}{\kappa t+1}a^{\dagger }a}H_{l,l}\left[ \frac{ia^{\dagger
}}{\sqrt{\kappa t\left( \kappa t+1\right) }},\frac{ia}{\sqrt{\kappa t\left(
\kappa t+1\right) }}\right] \colon .  \label{26}
\end{eqnarray}%
Using the definition of Laguerre-polynomial 
\begin{equation}
L_{l}\left( x\right) =\sum \binom{l}{l-k}\frac{\left( -x\right) ^{k}}{k!}
\label{27}
\end{equation}%
and%
\begin{equation}
L_{l}\left( xy\right) =\frac{\left( -1\right) ^{l}}{l!}H_{l,l}(x,y),
\label{28}
\end{equation}%
then (\ref{26}) becomes%
\begin{equation}
\rho \left( t\right) =\frac{\left( \kappa t\right) ^{l}}{\left( \kappa
t+1\right) ^{l+1}}\colon L_{l}\left( \frac{-a^{\dagger }a}{\kappa t\left(
\kappa t+1\right) }\right) e^{\frac{-1}{\kappa t+1}a^{\dagger }a}\colon .
\label{29}
\end{equation}%
Noting $e^{\frac{-1}{\kappa t+1}a^{\dagger }a}$ represents a chaotic photon
field, so $\rho \left( t\right) $ is a Laguerre-polynomial-weighted chaotic
field. Thus we see $\left\vert l\right\rangle \left\langle l\right\vert $
evolves into the mixed state (\ref{29}), so this diffusion process
manifestly embodies quantum decoherence.

As Eq. (\ref{29}) is just in the type of Eq. (\ref{2}), so we can confirm
the state described by Eq. (\ref{2}) indeed exists as an quantum optical
field.

Before we check $Tr\rho \left( t\right) =1$ for Eq. (29)$,$ let us present
an integration formula%
\begin{equation}
\int \frac{d^{2}\alpha }{\pi }e^{\lambda |\alpha |^{2}}|\alpha |^{2k}=\left( 
\frac{\partial }{\partial \lambda }\right) ^{k}\int \frac{d^{2}\alpha }{\pi }%
e^{\lambda |\alpha |^{2}}=k!\left( \frac{-1}{\lambda }\right) ^{k+1},
\label{30}
\end{equation}%
then we introduce the completeness relation of coherent state%
\begin{equation}
\int \frac{d^{2}\alpha }{\pi }\left\vert \alpha \right\rangle \left\langle
\alpha \right\vert =1  \label{31}
\end{equation}%
here $\left\vert \alpha \right\rangle =\exp \left[ \alpha a^{\dagger
}-|\alpha |^{2}/2\right] \left\vert 0\right\rangle $. Due to 
\begin{equation}
a\left\vert \alpha \right\rangle =\alpha \left\vert \alpha \right\rangle
,\left\langle \alpha \right\vert \colon f\left( a^{\dagger },a\right) \colon
\left\vert \alpha \right\rangle =f\left( \alpha ^{\ast },\alpha \right) ,
\label{32}
\end{equation}%
we see%
\begin{eqnarray}
&&\left\langle \alpha \right\vert \colon L_{l}\left( \frac{-a^{\dagger }a}{%
\kappa t\left( \kappa t+1\right) }\right) e^{\frac{-1}{\kappa t+1}a^{\dagger
}a}\colon \left\vert \alpha \right\rangle  \notag \\
&=&e^{\frac{-1}{\kappa t+1}|\alpha |^{2}}\sum_{k=0}^{l}\frac{l!(-1)^{k}}{k!%
\left[ (l-k)!\right] ^{2}}\left( \frac{-|\alpha |^{2}}{\kappa t\left( \kappa
t+1\right) }\right) ^{l-k}.  \label{33}
\end{eqnarray}%
Substituting (\ref{33}) into $tr\rho \left( t\right) =\int \frac{d^{2}\alpha 
}{\pi }\left\langle \alpha \right\vert \rho \left( t\right) \left\vert
\alpha \right\rangle $ we should calculate%
\begin{eqnarray}
tr\rho \left( t\right) &=&\int \frac{d^{2}\alpha }{\pi }\left\langle \alpha
\right\vert \rho \left( t\right) \left\vert \alpha \right\rangle  \notag \\
&=&\int \frac{d^{2}\alpha }{\pi }\left\langle \alpha \right\vert \frac{%
\left( \kappa t\right) ^{l}}{\left( \kappa t+1\right) ^{l+1}}\colon e^{\frac{%
-1}{\kappa t+1}a^{\dagger }a}L_{l}\left( \frac{-a^{\dagger }a}{\kappa
t\left( \kappa t+1\right) }\right) \colon \left\vert \alpha \right\rangle 
\notag \\
&=&\frac{\left( \kappa t\right) ^{l}}{\left( \kappa t+1\right) ^{l+1}}%
\sum_{k=0}^{l}\frac{l!}{k!\left[ (l-k)!\right] ^{2}}\int \frac{d^{2}\alpha }{%
\pi }e^{\frac{-1}{\kappa t+1}\left\vert \alpha \right\vert ^{2}}\left( \frac{%
|\alpha |^{2}}{\kappa t\left( \kappa t+1\right) }\right) ^{l-k}.  \label{34}
\end{eqnarray}%
By setting $\frac{\alpha }{\sqrt{\kappa t\left( \kappa t+1\right) }}=\alpha
^{\prime }$ we reform the integration as%
\begin{eqnarray}
&&\int \frac{d^{2}\alpha }{\pi }e^{\frac{-1}{\kappa t+1}\left\vert \alpha
\right\vert ^{2}}\left( \frac{|\alpha |^{2}}{\kappa t\left( \kappa
t+1\right) }\right) ^{l-k}  \notag \\
&=&\kappa t\left( \kappa t+1\right) \int \frac{d^{2}\alpha ^{\prime }}{\pi }%
e^{-\kappa t\left\vert \alpha ^{\prime }\right\vert ^{2}}\left( |\alpha
^{\prime }|^{2}\right) ^{l-k}  \notag \\
&=&\kappa t\left( \kappa t+1\right) \left( l-k\right) !\left( \frac{1}{%
\kappa t}\right) ^{l-k+1}.  \label{35}
\end{eqnarray}%
Substituting (\ref{35}) into (\ref{34}) we see%
\begin{equation}
tr\rho \left( t\right) =\frac{\left( \kappa t\right) ^{l+1}}{\left( \kappa
t+1\right) ^{l}}\sum_{k=0}^{l}\frac{l!}{k!(l-k)!}\left( \frac{1}{\kappa t}%
\right) ^{l-k+1}=1  \label{36}
\end{equation}%
so it is trace conservative.

\section{The photon number in the mixed state}

Then we evaluate the photon number for Eq. $(29)$%
\begin{eqnarray}
Tr\left[ \rho \left( t\right) a^{\dagger }a\right]  &=&Tr\left[ \rho \left(
t\right) aa^{\dagger }\right] -1  \notag \\
&=&\frac{\left( \kappa t\right) ^{l}}{\left( \kappa t+1\right) ^{l+1}}Tr%
\left[ \colon a^{\dagger }ae^{\frac{-1}{\kappa t+1}a^{\dagger }a}L_{l}\left( 
\frac{-a^{\dagger }a}{\kappa t\left( \kappa t+1\right) }\right) \colon %
\right] -1.  \label{37}
\end{eqnarray}%
By using the coherent state representation we have%
\begin{eqnarray}
&&Tr\left[ \colon a^{\dagger }ae^{\frac{-1}{\kappa t+1}a^{\dagger
}a}L_{l}\left( \frac{-a^{\dagger }a}{\kappa t\left( \kappa t+1\right) }%
\right) \colon \right]   \notag \\
&=&\int \frac{d^{2}\alpha }{\pi }e^{\frac{-1}{\kappa t+1}\left\vert \alpha
\right\vert ^{2}}\left\vert \alpha \right\vert ^{2}L_{l}\left( \frac{%
-\left\vert \alpha \right\vert ^{2}}{\kappa t\left( \kappa t+1\right) }%
\right)   \notag \\
&=&\left[ \kappa t\left( \kappa t+1\right) \right] ^{2}\int \frac{%
d^{2}\alpha ^{\prime }}{\pi }e^{-\kappa t\left\vert \alpha ^{\prime
}\right\vert ^{2}}L_{l}\left( -|\alpha ^{\prime }|^{2}\right) \left\vert
\alpha ^{\prime }\right\vert ^{2}  \notag \\
&=&\left[ \kappa t\left( \kappa t+1\right) \right] ^{2}\sum_{k=0}^{l}\frac{l!%
}{k!k!(l-k)!}\int \frac{d^{2}\alpha ^{\prime }}{\pi }e^{-\kappa t\left\vert
\alpha ^{\prime }\right\vert ^{2}}\left( |\alpha ^{\prime }|^{2}\right)
^{k+1}  \notag \\
&=&\left( \kappa t+1\right) ^{2}\sum_{k=0}^{l}\frac{l!}{k!(l-k)!}\left(
k+1\right) \left( \frac{1}{\kappa t}\right) ^{k},  \label{38}
\end{eqnarray}%
\newline
where%
\begin{eqnarray}
&&\sum_{k=0}^{l}\frac{l!}{k!(l-k)!}k\left( \frac{1}{\kappa t}\right) ^{k} 
\notag \\
&=&\frac{l}{\kappa t}\sum_{k=1}^{l}\frac{\left( l-1\right) !}{\left(
k-1\right) !(l-k)!}\left( \frac{1}{\kappa t}\right) ^{k-1}=\frac{l}{\kappa t}%
\left( \frac{\kappa t+1}{\kappa t}\right) ^{l-1},  \label{39}
\end{eqnarray}%
and%
\begin{equation}
\sum_{k=0}^{l}\frac{l!}{k!(l-k)!}\left( \frac{1}{\kappa t}\right)
^{k}=\left( \frac{\kappa t+1}{\kappa t}\right) ^{l}.  \label{40}
\end{equation}%
Substituting (\ref{39})-(\ref{40}) into (\ref{38}) we obtain%
\begin{eqnarray}
&&Tr\left[ \colon a^{\dagger }ae^{\frac{-1}{\kappa t+1}a^{\dagger
}a}L_{l}\left( \frac{-a^{\dagger }a}{\kappa t\left( \kappa t+1\right) }%
\right) \colon \right]   \notag \\
&=&\left( \kappa t+1\right) ^{2}\left( \frac{\kappa t+1}{\kappa t}\right)
^{l-1}\frac{l+\kappa t+1}{\kappa t}.  \label{41}
\end{eqnarray}%
Then substituting (\ref{41}) into (\ref{37}) we see%
\begin{equation}
Tr\left[ \rho \left( t\right) a^{\dagger }a\right] =l+\kappa t.  \label{42}
\end{equation}%
which tells that the photon number $l\rightarrow l+\kappa t.$

At the end of this work we point out that a diffusion process is quite
different from the process in the amplitude dissipative channel (ADC)
described by the following master equation [7] 
\begin{equation}
\frac{d}{dt}\rho ^{\prime }=\gamma \left( 2a\rho ^{\prime }a^{\dag }-a^{\dag
}a\rho ^{\prime }-\rho ^{\prime }a^{\dag }a\right)   \label{43}
\end{equation}%
where $\gamma $ is the rate of dissipation. The solution to Eq. (\ref{43})
is [8] 
\begin{equation}
\rho ^{\prime }=\sum_{m=0}^{\infty }\frac{\left( 1-e^{-2\gamma t}\right) ^{n}%
}{n!}e^{-\gamma ta^{\dag }a}a^{n}\rho _{0}^{\prime }a^{\dag n}e^{-\gamma
ta^{\dag }a}.  \label{44}
\end{equation}%
In ADC an initial pure number state $\left\vert l\right\rangle \left\langle
l\right\vert $ will evolve into a binomial state as shown in [9]%
\begin{equation}
\sum_{m=0}^{l}\binom{l}{l-m}e^{-2\gamma mt}\left( 1-e^{-2\gamma t}\right)
^{l-m}\left\vert m\right\rangle \left\langle m\right\vert \equiv \rho
_{b}^{\prime }  \label{45}
\end{equation}%
with photon number decaying $tr\left( a^{\dagger }a\rho _{b}^{\prime
}\right) =le^{-2\gamma t}$. Comparing the diffusion master equation $(3)$
with the dissipation Eq. $(43)$ we realize that the term $a^{\dagger }\rho a$
may be responsible for diffusion.

In summary, we theoretically propose a new optical field state%
\begin{equation}
\rho _{new}=\lambda \left( 1-\lambda \right) ^{l}\colon L_{k}\left( \frac{%
-\lambda ^{2}a^{\dag }a}{1-\lambda }\right) e^{-\lambda a^{\dag }a}\colon
\label{46}
\end{equation}%
which is named Laguerre-polynomial-weighted chaotic field. We show that such
state can be implemented, i.e., when a number state enters into a diffusion
channel, the output state is just this kind of states. We solve the master
equation describing the diffusion process by using the summation method
within ordered product of operators and the entangled state representaion.
The solution manifestly shows how a pure state evolves into a mixed state.
The physical difference between the diffusion and the amplitude damping is
pointed out.

\textbf{Acknowledgements:} This work was supported by the National Natural
Science Foundation of China (Grant Nos. 11175113 and 11264018), and the
Natural Science Foundation of Jiangxi Province of China (No 20132BAB212006).

\textbf{Appendix 1}

For $\rho _{new}$ in Eq. (\ref{2}) we prove $tr\rho _{new}=1.$ In fact,
using the coherent state representation we have%
\begin{align}
tr\rho _{new}& =\int \frac{d^{2}\alpha }{\pi }\left \langle \alpha \right
\vert \lambda \left( 1-\lambda \right) ^{l}\colon L_{l}\left( \frac{-\lambda
^{2}a^{\dag }a}{1-\lambda }\right) e^{-\lambda a^{\dag }a}\colon \left \vert
\alpha \right \rangle  \notag \\
& =\lambda \left( 1-\lambda \right) ^{l}\int \frac{d^{2}\alpha }{\pi }%
e^{-\lambda |\alpha |^{2}}L_{l}\left( \frac{-\lambda ^{2}|\alpha |^{2}}{%
1-\lambda }\right)  \notag \\
& =\lambda \left( 1-\lambda \right) ^{l}\int_{0}^{\infty }d\left( \frac{%
\lambda -1}{\lambda ^{2}}x\right) e^{-\frac{\lambda -1}{\lambda }%
x}L_{l}\left( x\right) =1,  \tag{A1}
\end{align}%
where we have used 
\begin{equation}
\int_{0}^{\infty }e^{-bx}L_{l}\left( x\right) =\left( b-1\right)
^{l}b^{-l-1}.  \tag{A2}
\end{equation}%
\textbf{Appendix 2 }

For $M_{m,n}$ in Eq. $\left( \ref{22}\right) $ we now prove $%
\sum_{m,n=0}^{\infty }M_{m,n}^{\dagger }M_{m,n}=1.$ Because%
\begin{equation}
\vdots e^{xaa^{\dagger }}\vdots =\frac{1}{1-x}e^{a^{\dagger }a\ln \frac{1}{%
1-x}},  \tag{A3}
\end{equation}%
where $\vdots $ $\vdots $ denotes anti-normal ordering, we have%
\begin{align}
& \sum_{m,n=0}^{\infty }\frac{1}{m!}\frac{\left( \kappa t\right) ^{m}}{%
\left( \kappa t+1\right) ^{m}}a^{m}a^{\dagger m}  \notag \\
& =\vdots \exp \left[ \frac{\kappa t}{\kappa t+1}aa^{\dagger }\right] \vdots
\notag \\
& =\left( \kappa t+1\right) e^{a^{\dagger }a\ln \left( \kappa t+1\right) }. 
\tag{A4}
\end{align}%
Substituing it into the sum representation of $\rho \left( t\right) $ yields%
\begin{align}
\sum_{m,n=0}^{\infty }M_{m,n}^{\dagger }M_{m,n}& =\sum_{m,n=0}^{\infty }%
\frac{1}{m!n!}\frac{\left( \kappa t\right) ^{m+n}}{\left( \kappa t+1\right)
^{m+n+1}}a^{\dagger n}\left( \frac{1}{1+\kappa t}\right) ^{a^{\dagger
}a}a^{m}a^{\dagger m}\left( \frac{1}{1+\kappa t}\right) ^{a^{\dagger }a}a^{n}
\notag \\
& =\sum_{n=0}^{\infty }\frac{1}{n!}\frac{\left( \kappa t\right) ^{n}}{\left(
\kappa t+1\right) ^{n}}a^{\dagger n}\left( \frac{1}{1+\kappa t}\right)
^{2a^{\dagger }a}e^{a^{\dagger }a\ln \left( \kappa t+1\right) }a^{n}  \notag
\\
& =\sum_{n=0}^{\infty }\frac{1}{n!}\frac{\left( \kappa t\right) ^{n}}{\left(
\kappa t+1\right) ^{n}}a^{\dagger n}e^{a^{\dagger }a\ln \left( \kappa
t+1\right) }e^{2a^{\dagger }a\ln \frac{1}{1+\kappa t}}a^{n}  \notag \\
& =\sum_{n=0}^{\infty }\frac{1}{n!}\frac{\left( \kappa t\right) ^{n}}{\left(
\kappa t+1\right) ^{n}}a^{\dagger n}e^{a^{\dagger }a\left[ \ln \left( \kappa
t+1\right) +2\ln \frac{1}{1+\kappa t}\right] }a^{n}  \notag \\
& =\sum_{n=0}^{\infty }\frac{1}{n!}\frac{\left( \kappa t\right) ^{n}}{\left(
\kappa t+1\right) ^{n}}a^{\dagger n}e^{a^{\dagger }a\ln \frac{1}{1+\kappa t}%
}a^{n}  \notag \\
& =\sum_{n=0}^{\infty }\frac{1}{n!}\frac{\left( \kappa t\right) ^{n}}{\left(
\kappa t+1\right) ^{n}}a^{\dagger n}\colon e^{a^{\dagger }a\left( \frac{1}{%
1+\kappa t}-1\right) }\colon a^{n}  \notag \\
& =\colon e^{a^{\dagger }a\frac{\kappa t}{1+\kappa t}}e^{a^{\dagger }a\ln 
\frac{-\kappa t}{1+\kappa t}}\colon =1.  \tag{A5}
\end{align}

\end{document}